\documentclass[prd,superscriptaddress,amsfonts,amssymb,amsmath,showpacs,onecolumn]{revtex4-2}
\usepackage{bm}
\usepackage{amsfonts}
\usepackage{latexsym}
\usepackage[latin1]{inputenc}
\usepackage{graphicx}
\usepackage{amsmath,eqnarray}
\usepackage{palatino}
\usepackage{mathpazo}
\usepackage{textcomp}
\usepackage{float}
\usepackage{booktabs}
\usepackage{dcolumn}
\usepackage{hyperref}
\usepackage{mathtools}
\hypersetup{colorlinks,citecolor=red}
\usepackage{ragged2e}
\usepackage{amsmath}
\usepackage{xcolor}
\usepackage{orcidlink}
\usepackage[caption=false]{subfig}
\usepackage{commath}
\def\jnl@style{\it}
\def\aaref@jnl#1{{\jnl@style#1}}

\def\aaref@jnl#1{{\jnl@style#1}}

\def\aj{\aaref@jnl{AJ}}                   
\def\apj{\aaref@jnl{ApJ}}                 
\def\apjl{\aaref@jnl{ApJ}}                
\def\apjs{\aaref@jnl{ApJS}}               
\def\apss{\aaref@jnl{Ap\&SS}}             
\def\aap{\aaref@jnl{A\&A}}                
\def\aapr{\aaref@jnl{A\&A~Rev.}}          
\def\aaps{\aaref@jnl{A\&AS}}              
\def\mnras{\aaref@jnl{Mon.~Not.~Roy.~Astron.~Soc.}}             
\def\prd{\aaref@jnl{Phys.~Rev.~D}}        
\def\prc{\aaref@jnl{Phys.~Rev.~C}}  
\def\prl{\aaref@jnl{Phys.~Rev.~Lett.}}    
\def\qjras{\aaref@jnl{QJRAS}}             
\def\skytel{\aaref@jnl{S\&T}}             
\def\ssr{\aaref@jnl{Space~Sci.~Rev.}}     
\def\zap{\aaref@jnl{ZAp}}                 
\def\nat{\aaref@jnl{Nature}}              
\def\aplett{\aaref@jnl{Astrophys.~Lett.}} 
\def\apspr{\aaref@jnl{Astrophys.~Space~Phys.~Res.}} 
\def\physrep{\aaref@jnl{Phys.~Rep.}}      
\def\physscr{\aaref@jnl{Phys.~Scr}}       
\def\commat{\aaref@jnl{Comm.~Math.~Phys.}}              
\def\science{\aaref@jnl{Science}}               
\def\cqg{\aaref@jnl{Classical Quant.~Grav.}}            
\def\jpcs{\aaref@jnl{JPCS}}                                     
\def\ijmpd{\aaref@jnl{Int.~J.~Mod.~Phys.~D}}                    
\def\grg{\aaref@jnl{Gen.~Relat.~Gravit.}}               
\def\rpp{\aaref@jnl{Rep.~Prog.~Phys.}}          
\def\npa{\aaref@jnl{Nucl.~Phys.~A}}        
\def\lrr{\aaref@jnl{Living Rev.~Rel.}}                   
\def\jcap{\aaref@jnl{J.~Cosmology Astropart.~Phys.}}    
\def\rmp{\aaref@jnl{Rev.~Mod.~Phys.}}   
\def\Eur. Phys. J. C.{\aaref@jnl{Eur.~Phys.~J.~C}}

\allowdisplaybreaks[1]

\begin{document}

\color{black}

\title{Cosmological constraints on $f(Q)$ gravity models in the non-coincident formalism}

\author{Sneha Pradhan\orcidlink{0000-0002-3223-4085}}
\email{snehapradhan2211@gmail.com}

\author{Raja Solanki\orcidlink{0000-0001-8849-7688}}
\email{rajasolanki8268@gmail.com}
\affiliation{Department of Mathematics, Birla Institute of Technology and
Science-Pilani,\\ Hyderabad Campus, Hyderabad-500078, India.}

\author{P.K. Sahoo\orcidlink{0000-0003-2130-8832}}
\email{pksahoo@hyderabad.bits-pilani.ac.in}
\affiliation{Department of Mathematics, Birla Institute of Technology and
Science-Pilani,\\ Hyderabad Campus, Hyderabad-500078, India.}

\date{\today}

\begin{abstract}
The article investigates cosmological applications of $f(Q)$ theories in a non-coincident formalism. We explore a new $f(Q)$ theory dynamics utilizing a non-vanishing affine connection involving a non-constant function $\gamma(t)=-a^{-1}\dot{H}$, resulting in Friedmann equations that are entirely distinct from those of $f(T)$ theory. In addition, we propose a new parameterization of the Hubble function that can consistently depicts the present deceleration parameter value, transition redshift, and the late time de-Sitter limit. We evaluate the predictions of the assumed Hubble function by imposing constraints on the free parameters utilizing Bayesian statistical analysis to estimate the posterior probability by employing the CC, Pantheon+SH0ES, and the BAO samples. Moreover, we conduct the AIC and BIC statistical evaluations to determine the reliability of MCMC analysis. Further, we consider some well-known corrections to the STEGR case such as an exponentital $f(Q)$ correction, logarithmic $f(Q)$ correction, and a power-law $f(Q)$ correction and then we find the constraints on the parameters of these models via energy conditions. Finally, to test the physical plausibility of the assumed $f(Q)$ models we conduct the thermodynamical stability analysis via the sound speed parameter.\\

\end{abstract}
\maketitle

\section{Introduction}\label{sec1}

Theoretical advancements and modifications in gravity theories \cite{a1} are pivotal in interpreting cosmological data \cite{a2}-\cite{a5}. Modifying the gravitational action integral by incorporating geometric invariants introduces new degrees of freedom into the gravitational field equations. This deviation from general relativity (GR) enables a more accurate theoretical alignment with observations \cite{a6}. Einstein's GR relies on the Ricci scalar, 
$R$, determined by the symmetric Levi-Civita connection. A fundamental alteration to the Einstein-Hilbert action involves incorporating a function $f$ of this scalar, resulting in the so-called  $f(R)$-theory of gravity \cite{a7}. In a gravitational theory, the Levi-Civita connection is not the only possible choice. By employing a more general connection, one can define the fundamental scalar invariants of curvature $R$, torsion $T$, and non-metricity $Q$. The nature of the physical space is determined by the invariant used to define the gravitational action integral \cite{a8}. Specifically, GR arises from the Levi-Civita connection, where only the curvature invariant is relevant. Alternatively, the curvature-less Weitzenbock connection \cite{a9} leads to the Teleparallel Equivalent of General Relativity \cite{a10}-\cite{a11}. Moreover, a gravitational theory based on a torsion-free, flat connection results in the Symmetric teleparallel equivalent of GR (TEGR) \cite{a12}. Although these three theories produce identical field equations, their modifications differ significantly. For instance, the modified theories, such as the 
$f(T)$ teleparallel theory \cite{a13} and the $f(Q)$ symmetric teleparallel theory \cite{a14}, are remarkably different from the $f(R)$ theory.

This research delves into identifying cosmological solutions in the $f(Q)$ symmetric teleparallel theory. There are a number of studies on the $f(Q)$ theory that are extensive. For instance, \cite{a15},\cite{a16} provides the exact analytic cosmological solutions by implementing a unique model. The researcher has examined in \cite{a17}  the power-law functions of 
$f(Q)$ and their parameter values. The potential of 
$f(Q)$ theory as a dark energy model is explored in \cite{a18}-\cite{a21}. Recent research into the effective fluid properties due to non-metricity is found in \cite{a22}. Anisotropic spacetimes are discussed in \cite{a23}-\cite{a27}. The Hamiltonian analysis of this theory is detailed in \cite{a29}, while the cosmological quantization process is presented in \cite{a30}.  Apart from that some wormhole solutions and the spherically symmetric solution of the compact object in the $f(Q)$ gravity context are given in \cite{a31}-\cite{b4}. Studies on non-minimal couplings to matter are found in \cite{a32}, while generalizations including the energy-momentum tensor trace are discussed in \cite{a33,a34}, also the observational constraints are explored in \cite{a35}-\cite{a36}.

 Upon a thorough review of the literature on cosmological applications of 
$f(Q)$theories, we observed a prevalent trend of utilizing only the vanishing affine connection in the spatially flat Friedmann-Lemaitre-Robertson-Walker (FLRW) spacetime. However, in this scenario, the Friedmann equations are identical to those of the 
$f(T)$ theory \cite{a15}. As a result, this specific gauge limits researchers to the outcomes already achieved in $f(T)$ theory, undermining the significance of $f(Q)$ theory as a novel modified gravity theory. A new framework \cite{ab2} for $f(Q)$ theory has been introduced, based on a different gauge equivalence class of non-vanishing affine connections, incorporating a function
$\gamma$ in the spatially flat FLRW background.
In this manuscript, we explore the observational constraints on several well known class of $f(Q)$ gravity models in this novel non-coincident formalism, along with a new parameterization of the Hubble function that can adequately describes the different cosmological epochs such as late time acceleration with recent transtion from deceleration epoch to acceleration epoch with observational compatibility. In section \ref{sec2}, we introduce fundamental formulations of the $f(Q)$ gravity and its covariant form. In section \ref{sec3}, we present Friedmann like equations for the $f(Q)$ gravity formalism in a non-coincident gauge setting. We also propose a new parameterization scheme of the Hubble function. In section \ref{sec4}, we carry out a statistical analysis to evaluate the predictions of the proposed Hubble function using recent observational data. Further in section \ref{sec5}, we test the viability of different $f(Q)$ gravity models utilizing the thermodynamical stability and produce the parameter constraints on these models with the help of energy conditions. In section \ref{sec6}, we conclude our outcomes of the investigation.

\section{Mathematical Formulation}\label{sec2}
In the symmetric teleparallel theory of gravity, the focus is on a general affine connection 
$\Gamma^{\lambda}{ }_{\mu \nu}$ 
that is characterized by zero curvature and zero torsion, allowing the non-metricity to solely govern the gravitational interactions. The non-metricity tensor is thus defined within this framework.
\begin{eqnarray}
Q_{\lambda \mu \nu}=\nabla_{\lambda} g_{\mu \nu}.
\end{eqnarray}

The non-metricity tensor has two distinct types of traces, defined as follows:
\begin{eqnarray}
Q_{\lambda} = Q_{\lambda \mu \nu} g^{\mu \nu}, \quad \tilde{Q}_{\nu} = Q_{\lambda \mu \nu} g^{\lambda \mu}.
\end{eqnarray}

The superpotential tensor \(P^{\lambda}{}_{\mu \nu}\) and disformation tensor \(L^{\lambda}{}_{\mu \nu}\) are expressed as follows:
\begin{eqnarray}
&&\hspace{0.0cm}P_{~~\mu \nu}^{\lambda} = \frac{1}{4} \big( -2 L^{\lambda}{}_{\mu \nu} + Q^{\lambda} g_{\mu \nu} - \tilde{Q}^{\lambda} g_{\mu \nu} - \frac{1}{2} \delta_{\mu}^{\lambda} Q_{\nu}\nonumber\\&&\hspace{1.5cm} - \frac{1}{2} \delta_{\nu}^{\lambda} Q_{\mu} \big). \label{3}
\end{eqnarray}
\begin{eqnarray}
&&\hspace{0.0cm}L_{\mu \nu}^{\lambda} = \frac{1}{2} \left( Q^{\lambda}{}_{\mu \nu} - Q_{\mu}{}^{\lambda}{}_{\nu} - Q_{\nu}{}^{\lambda}{}_{\mu} \right), \label{2}
\end{eqnarray}

It is widely known that the disformation tensor establishes the relationship between the affine connection and the Levi-Civita connection $\Hat{\Gamma}{}^{\lambda}{}_{\mu \nu}$ as follows:

\begin{eqnarray}
&&\hspace{0.2cm}\Gamma^{\lambda}{}_{\mu \nu} = \Hat{\Gamma}{}^{\lambda}{}_{\mu \nu}+ L^{\lambda}{}_{\mu \nu}.
\footnotetext{For a comprehensive study on the application of autonomous dynamical systems in various modified theories of gravity, see Refs. [10-14] and also Ref. [43].}
\end{eqnarray}

The non-metricity scalar can be defined as:
\begin{eqnarray}
   && \hspace{0.0cm} Q = Q_{\lambda \mu \nu} P^{\lambda \mu \nu} \nonumber\\&& \hspace{0.5cm}= \frac{1}{4} \left( -Q_{\lambda \mu \nu} Q^{\lambda \mu \nu} + 2 Q_{\lambda \mu \nu} Q^{\mu \lambda \nu} + Q_{\lambda} Q^{\lambda} - 2 Q_{\lambda} \tilde{Q}^{\lambda} \right).~~
\end{eqnarray}

Although symmetric teleparallelism is equivalent to GR, it shares the same \textit{dark} issues inherent in GR. To address these issues, a modified \(f(Q)\) gravity theory has been proposed \cite{a14}, analogous to the \(f(R)\) extension of GR or $f(T)$ extension of teleparallel gravity. The Einstein-Hilbert action in \(f(Q)\) theory under the symmetric teleparallel framework is given by,
\begin{eqnarray}
S = \frac{1}{2 \kappa} \int f(Q) \sqrt{-g} \, d^{4}x + \int \mathcal{L}_{M} \sqrt{-g} \, d^{4}x,
\end{eqnarray}
Now by varying the action with respect to the metric inverse $g^{\mu\nu}$, we obtain the field equation:
\begin{eqnarray}\label{fe11}
  &&\hspace{-0.0cm}\frac{2}{\sqrt{-g}} \nabla_{\lambda} ( \sqrt{-g} F P_{~~~\mu \nu}^{\lambda} ) - \frac{1}{2} f g_{\mu \nu} + F \big( P_{\nu \rho \sigma} Q_{\mu}^{~~\rho \sigma} \nonumber\\&&\hspace{0.6cm}- 2 P_{\rho \sigma \mu} Q_{~~\nu}^{\rho \sigma} \big) = \kappa T_{\mu \nu}^{m},~~~~~~~~~~~~~~~~~~~~~~~~~~~~~~~~~~~~   
\end{eqnarray}

where \(F = \frac{d f}{d Q}\) (with derivatives indicated by primes hereafter). The stress-energy tensor \(T_{~~\mu \nu}^{m}\) is assumed to be that of a perfect fluid:
\begin{eqnarray}
T_{~~\mu \nu}^{m} = (p + \rho) u_{\mu} u_{\nu} + p g_{\mu \nu},
\end{eqnarray}
where \(p\) and \(\rho\) represent the pressure and energy density of the matter.

Recently, the covariant formulation of this field equation has been developed and effectively applied in the cosmological context \cite{cov1,cov2}:
\begin{eqnarray}
F \Hat{G}_{\mu\nu} + \frac{1}{2} g_{\mu \nu} (F Q - f) + 2 F' P^{\lambda}{}_{\mu \nu} \Hat{\nabla}_{\lambda} Q = \kappa T_{~\mu \nu}^{m},
\end{eqnarray}
where
\begin{eqnarray}
\Hat{G}_{\mu\nu} = \Hat{R}_{\mu\nu} - \frac{1}{2} g_{\mu \nu} \Hat{R}.
\end{eqnarray}
All expressions with an upper cap are calculated concerning the Levi-Civita connection. Thus, in its GR equivalent form, we can express it as :
\begin{eqnarray}
\Hat{G}_{\mu\nu} = \frac{\kappa}{F} \mathcal{T}_{\mu \nu} = \frac{\kappa}{F} T_{\mu \nu}^{m} + \kappa T_{\mu \nu}^{\text{DE}},
\end{eqnarray}
where
\begin{eqnarray}
\kappa T_{~~\mu \nu}^{\text{DE}} = \frac{1}{F} \left[ \frac{1}{2} g_{\mu \nu} (f - Q F) - 2 F' \Hat{\nabla}_{\lambda} Q P_{\mu \nu}^{\lambda} \right].
\end{eqnarray}

Furthermore, By varying the action with respect to the affine connection, and assuming that the matter Lagrangian \(\mathcal{L}_{M}\) is independent of the affine connection, we derive another field equation for the \(f(Q)\) theory:

\begin{eqnarray}\label{field}
\nabla_{\mu} \nabla_{\nu} \left( \sqrt{-g} F P_{\lambda}{}^{\nu \mu} \right) = 0.
\end{eqnarray}

\section{Equations of Motion}\label{sec3}
The Friedmann-Lemaitre-Robertson-Walker (FLRW) metric for  spatially flat homogeneous and isotropic case, expressed in spherical coordinates, is given by:

\begin{eqnarray}\label{metric}
dS^2= & -\mathrm{d} t \otimes \mathrm{d} t+a(t)^{2}\left(\mathrm{~d} r \otimes \mathrm{d} r+r^{2} \mathrm{~d} \theta \otimes \mathrm{d} \theta\right. \left.+r^{2} \sin^{2}\theta \mathrm{~d} \phi \otimes \mathrm{d} \phi\right) 
\end{eqnarray}

Several significant publications have recently emerged on the modified 
$f(Q)$ gravity theory and its cosmological implications, in the references \cite{2}-\cite{15}. Additionally, the dynamical system analysis of the $f(Q)$ theory at both background and perturbative levels for a spatially flat FLRW spacetime has been conducted in \cite{16}. These analyses, however, have predominantly utilized the coincident gauge choice. In this gauge, the line element is expressed in cartesian coordinates with $\Gamma^{\lambda}{}_{\mu \nu} = 0$. This simplification reduces the covariant derivative to a partial derivative, facilitating easier calculations.




In the present discussion, we examine a significant class of affine connections, denoted by $\Gamma$, which were initially introduced and analyzed from a cosmological standpoint in \cite{ab2}. These connections are torsion-free and have zero curvature, but they are not compatible with the metric given in equation (\ref{metric}). The non-vanishing components of the affine connection $\Gamma$ corresponding to the metric (\ref{metric}) are as follows,

\begin{eqnarray}
\nonumber
  &&\hspace{0cm}   \Gamma^{t}{ }_{t t}=\gamma+\frac{\dot{\gamma}}{\gamma}, \quad \Gamma^{r}{ }_{t r}=\gamma, \quad \Gamma^{r}{ }_{\theta \theta}=-r \\ \nonumber
&&\hspace{0cm} \Gamma^{r}{ }_{\phi \phi}=-r \sin ^{2} \theta, \quad \Gamma^{\theta}{ }_{t \theta}=\gamma \\ \nonumber
&&\hspace{0cm} \Gamma^{\theta}{ }_{r \theta}=\frac{1}{r}, \quad \Gamma^{\theta}{ }_{\phi \phi}=-\cos \theta \sin \theta, \quad \Gamma^{\phi}{ }_{t \phi}=\gamma \\
 &&\hspace{0cm} \Gamma^{\phi}{ }_{r \phi}=\frac{1}{r}, \quad \Gamma^{\phi}{ }_{\theta \phi}=\cot \theta
\end{eqnarray}

We can then calculate the required tensors in the general setting :

\begin{eqnarray}
&&\hspace{0cm}Q_{t t t}=2 \big(\gamma+\frac{\dot{\gamma}}{\gamma}\big), \quad Q_{t r}{ }^r=Q_{t \theta}{ }^\theta=Q_{t \phi}{ }^\phi=2\left(H-\gamma\right), ~~~~\\
&&\hspace{0cm} \quad Q^r{ }_{r t}=Q^\theta{ }_{\theta t}=Q^\phi{ }_{\phi t}=-\gamma, \\
&&\hspace{0cm} \quad L_{t t t}=-\big(\gamma+\frac{\dot{\gamma}}{\gamma}\big), \quad L_{t r}{ }^r=L_{t \theta}{ }^\theta=L_{t \phi}{ }^\phi=H,\\&&\hspace{0cm} \quad L_{r t}^r=L^\theta{ }_{\theta t}=L^\phi{ }_{\phi t}=\gamma-H, \\
&&\hspace{0cm}\quad P_{t t t}=\frac{-3\gamma}{4}, \quad P_{t r}{ }^r=P_{t \theta}{ }^\theta=P_{t \phi}{ }^\phi=\frac{1}{4}\left(4 H-3 \gamma\right),\\
&&\hspace{0cm}\quad P_{r t}^r=P^\theta{ }_{\theta t}=p^\phi{ }_{\phi t}=\frac{1}{4}\left(2\gamma+\frac{\dot{\gamma}}{\gamma}-H\right) .
\end{eqnarray}

In the above expression, $\gamma$ represents the nonvanishing function of time (t). An overdot denotes a time derivative. This function generates nonzero components in the nonmetricity tensor, offering new insights into the dynamics of $f(Q)$ in a spatially flat FLRW background, distinctly separate from the dynamics of $f(T)$ theory. However, due to the connection field equation \ref{field}, we face significant limitations on the permissible forms of the function $f$ \cite{AD}. The non-metricity scalar $Q$ can be calculated as follows:
\begin{eqnarray}
    Q=-6 H^{2}+9 \gamma H+3 \dot{\gamma} 
\end{eqnarray}
The equations analogous to the Friedmann equations, derived from the field equation \ref{fe11}, are as follows:

\begin{eqnarray}\label{f1}
  &&\hspace{0.5cm}  \kappa \rho=\frac{1}{2} f+\left(3 H^{2}-\frac{Q}{2}\right) F+\frac{3}{2} \dot{Q} \gamma F^{\prime},\\ \label{f2}
&&\hspace{0.0cm}\kappa p=-\frac{1}{2} f+\left(-2 \dot{H}-3 H^{2}+\frac{Q}{2}\right) F+\frac{\dot{Q}}{2}(-4 H+3 \gamma) F^{\prime}.~~~~~~~
\end{eqnarray}
In particular, for the choice $\gamma=0$, one can retrieve the usual Friedmann-like equation of the $f(Q)$ gravity in a \textit{coincident gauge} formalism. 

For further analysis, we propose a new parameterization Hubble function as follows,
\begin{equation}\label{NH}
H(z) = H_0 (z+1)^n + \beta \left[1-(z+1)^n\right].   
\end{equation}
Here $H_0$, $\beta$, and $n$ are free parameters. Several parameterizations have been explored for the equation of state (EoS) parameter, including the CPL (Chevallier-Polarski-Linder), BA (Barboza-Alcaniz), and LC (Low Correlation) models , as well as the deceleration parameter \cite{2,p1}. Beyond the parameterization of the deceleration parameter, numerous other schemes for parameterizing different cosmological parameters have been extensively discussed in the literature\cite{p2,p3}. These schemes aim to address issues in cosmological investigations, such as the initial singularity problem, the persistent accelerating expansion problem, the horizon problem, and the Hubble tension. However, several previous parameterization schemes \cite{sib1,sib2,sib3} leads to a discrepancy from the standard $\Lambda$CDM and hence cannot consistently describe the present deceleration parameter value, transition redshift, and the late time de-Sitter limit as obtained in standard $\Lambda$CDM model. This motivated us to consider a new parameterization function that can describe the aforementioned cosmological epochs with observational compatibility.

\section{Data and Methodology}\label{sec4}

In this section, we carry out a statistical analysis to evaluate how the predictions of the modified gravity models align with cosmological observational data. Our goal is to impose constraints on the free parameters. We use a dataset from Cosmic Chronometers, containing 31 measurements, and the Pantheon+SH0ES sample, which includes 1701 data points, as well as Baryonic Acoustic Oscillation data with 6 data points for our analysis.
\subsection{Statistical Methodology}
We utilize Bayesian statistical analysis to estimate the posterior probability by employing the likelihood function and the Markov Chain Monte Carlo (MCMC) random sampling technique. MCMC is widely used in cosmology to investigate the parameter space and generate corresponding probability distributions \cite{h1}. The core idea of MCMC is to create a Markov chain that samples a model's parameter space based on a specified probability distribution. This chain consists of a sequence of parameter values, where each value is derived from the previous one through a defined set of transition rules associated with a proposal distribution. The proposal distribution suggests a new parameter value, and its acceptance depends on its posterior probability, considering both the observational data and the prior probability function \cite{h2}. After the chain converges, the posterior distribution for the parameters can be estimated by calculating the frequency of parameter values within the chain. This posterior distribution then enables the estimation of optimal parameter values and their associated uncertainties, aiding in predictions for various observables.

\subsection{Cosmic Chronometers}
Cosmic chronometers are identified as a group of primarily old, passive galaxies that have ceased star formation, distinguishable by unique features in their spectra and color profiles \cite{h3}. The data used in cosmic chronometers is estimated by determining the age of these galaxies at various redshifts. In our study, we utilized a dataset of 31 independent measurements of \( H(z) \) within the redshift range \( 0.07 \leq z \leq 2.41 \) \cite{a3}. These \( H(z) \) measurements are derived from the relationship \( H(z) = - \frac{1}{1+z} \frac{dz}{dt} \). Here, \( \frac{dz}{dt} \) is estimated using \( \frac{\Delta z}{\Delta t} \), where \( \Delta z \) and \( \Delta t \) represent the change in redshift and age between two galaxies. The corresponding \( \chi^2 \) function is defined as:

\begin{eqnarray}
    \chi_H^2=\sum_{k=1}^{31} \frac{\left[H_{t h}\big(z_k, \theta\big)-H_{o b s}\left(z_k\right)\right]^2}{\sigma^2_{H(z_k)}}
\end{eqnarray}

Here, $H_{obs}(z_k)$ is the Hubble parameter value derived from cosmic observations, while $H_{th}(z,\theta)$
 represents its theoretical value calculated at redshift $z_k$ 
 with parameter space 
$\theta$. Additionally, $\sigma_{H(z_k)}$ denotes the associated error.

\subsection{Pantheon+SH0ES}
\justifying
The Pantheon+SH0ES samples span a wide range of redshifts, from 0.001 to 2.3, surpassing earlier collections of Type Ia supernovae (SNIa) by incorporating the latest observational data. Type Ia supernovae, known for their consistent brightness, serve as dependable standard candles for measuring relative distances using the distance modulus technique. Over the past two decades, several compilations of Type Ia supernova data have emerged, including Union \cite{a3}, Union2 \cite{R38}, Union2.1 \cite{R39}, JLA \cite{R40}, Pantheon \cite{R41}, and the most recent addition, Pantheon+SH0ES \cite{R42}. The corresponding $\chi^2$ function is given as,\begin{equation}\label{4b}
\chi^2_{SN}= D^T C^{-1}_{SN} D,
\end{equation}
$C_{SN}$ \cite{R42} denotes the covariance matrix related to the Pantheon+SH0ES samples, encompassing both statistical and systematic uncertainties. Additionally, the vector $D$ is defined as $D=m_{Bi}-M-\mu^{th}(z_i)$, where $m_{Bi}$ and $M$ denote the apparent magnitude and absolute magnitude, respectively. Furthermore, $\mu^{th}(z_i)$ represents the distance modulus of the assumed theoretical model, which can be expressed as,
\begin{equation}\label{4c}
\mu^{th}(z_i)= 5log_{10} \left[ \frac{D_{L}(z_i)}{1 Mpc}  \right]+25, 
\end{equation}

here $D_{L}(z)$ represents the luminosity distance in the proposed theoretical model, and it can be formulated as:
\begin{equation}\label{4d}
D_{L}(z)= c(1+z) \int_{0}^{z} \frac{ dx}{H(x,\theta)}
\end{equation}
here, $\theta$ denotes the parameter space of our constructed model.

The Pantheon+SH0ES compilation resolves the degeneracy between the parameters $H_0$ and $M$ differently from the Pantheon dataset by redefining the vector $D$ as follows:
\begin{equation}\label{4e}
\bar{D} = \begin{cases}
     m_{Bi}-M-\mu_i^{Ceph} & i \in \text{Cepheid hosts} \\
     m_{Bi}-M-\mu^{th}(z_i) & \text{otherwise}
    \end{cases}   
\end{equation}
Where, $\mu_i^{Ceph}$ is independently estimated using Cepheid calibrators. Consequently, the equation \eqref{4b} becomes $\chi^2_{SN} = \bar{D}^T C^{-1}_{SN} \bar{D}$.

\subsection{Baryonic Acoustic Oscillations}
\justifying
Baryonic Acoustic Oscillation (BAO) examines oscillations that originated in the early universe as a result of cosmological perturbations in a fluid composed of photons, baryons, and dark matter, all of which are closely linked through Thompson scattering. The BAO data include the Sloan Digital Sky Survey (SDSS), Six Degree Field Galaxy Survey (6dFGS), and the Baryon Oscillation Spectroscopic Survey (BOSS) \cite{58,59}. The equations utilized in BAO measurements are as follows.
$$
\begin{gathered}
d_A(z)=\int_0^z \frac{d z^{\prime}}{H\left(z^{\prime}\right)}, \\
D_V(z)=\left(d_A(z)^2 z / H(z)\right)^{1 / 3},
\end{gathered}
$$
and
$$
\chi_{B A O}^2=X^T C^{-1} X
$$
Here $C$ is known as the covariance matrix \cite{60}, $d_A(z)$ represents the angular diameter distance, while $D_V(z)$ stands for the dilation scale.

\subsection{CC + Type Ia Supernovae Sample+BAO}
\begin{widetext}
\begin{figure}[H]
\includegraphics[scale=0.85]{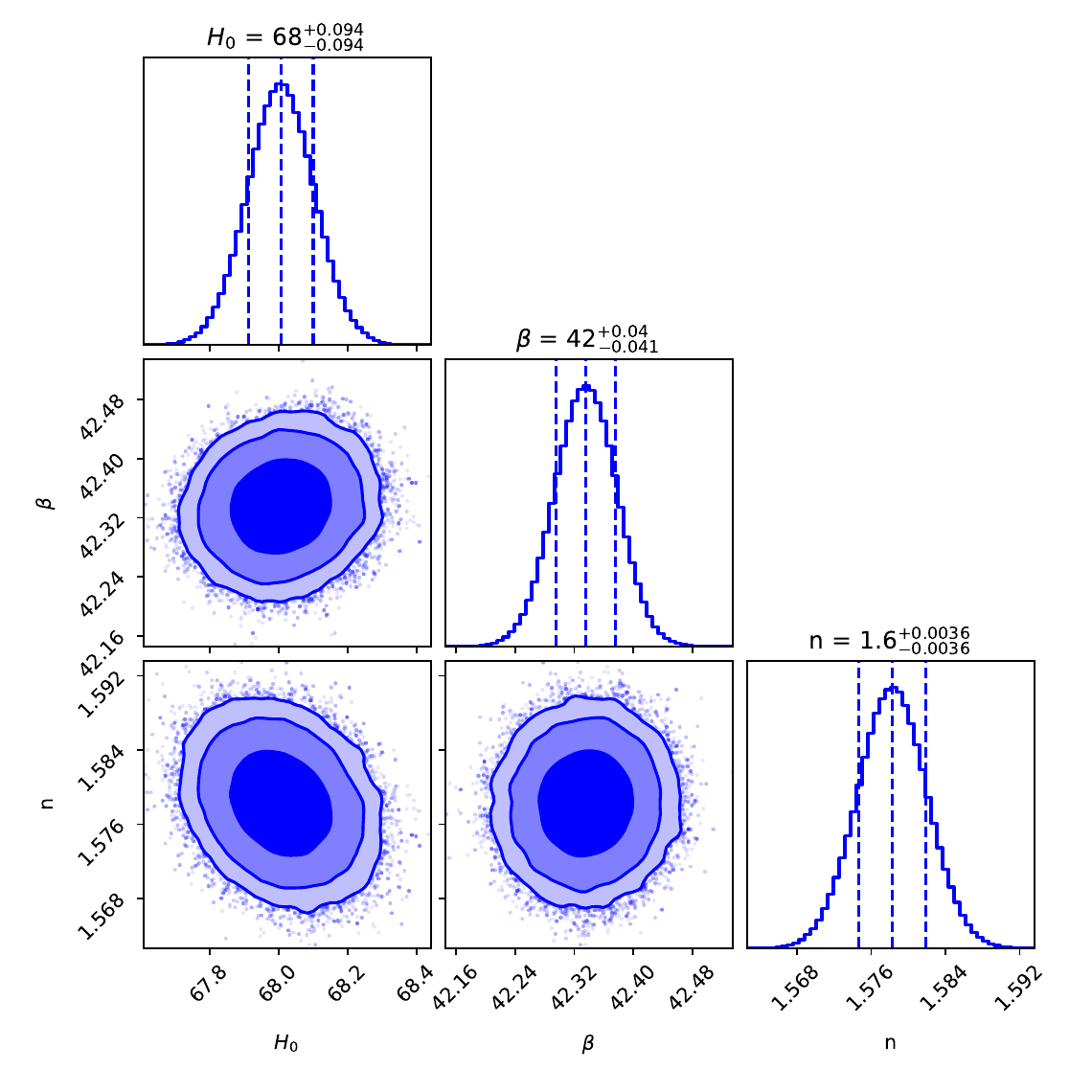}
\caption{The contour plot for the free parameter space $(H_0, \beta, n)$ corresponding to the given Hubble function within the $1\sigma-3\sigma$ confidence interval using CC+Pantheon+SH0ES+BAO samples.}\label{fig1} 
\end{figure} 
\end{widetext}

\justifying
We determine the constraints on the free parameter space for the combined CC+Pantheon+SH0ES+BAO samples using the Gaussian priors within the range of $H_0 \in [50,100]$, $\beta \in [30, 60]$, and $n \in [0,5]$. To find the best fit values for the parameters, we minimize the total $\chi^2_{total}$ function, defined as follows:
\begin{equation}\label{4f}
\chi^2_{total}= \chi^2_{CC}+\chi^2_{SN} +\chi^2_{BAO} 
\end{equation}
The contour plot describing the correlation between the free parameters within the $1\sigma-3\sigma$ confidence interval is presented in Fig.~\ref{fig1}.

We obtained the free parameter constraints as $H_0=68 \pm 0.094 \: km/s/Mpc$, $\beta=42^{+0.04}_{-0.041}$, and $n=1.6 \pm 0.0036$ within $68 \%$ confidence limit. In addition, we obtained the minimum value of the $\chi^2_{total}$ as $\chi^2_{min}=1648.988$.

\subsection{Model Comparison with $\Lambda$CDM}
\justifying
To determine the reliability of the MCMC analysis, it is essential to conduct a statistical evaluation using the Akaike Information Criterion (AIC) and Bayesian Information Criterion (BIC). The formulation of AIC could be written as \cite{chi}:
\begin{eqnarray}
    AIC=\chi^2_{min}+2d
\end{eqnarray}

Here, d denotes the number of parameters in the specified model. When comparing the model with the established $\Lambda$CDM model, we are establishing $\Delta \text{AIC}=|\text{AIC}_{\text{Model}}-\text{AIC}_{ \Lambda\text{CDM}}|$. A value in $\Delta$AIC of less than 2 indicates strong support for the assumed theoretical model. A value in $\Delta$AIC between 4 and 7 suggests moderate support. If the value of $\Delta$AIC exceeds 10, there is no evidence to support the assumed model. The second criterion, BIC, is formulated in the following manner:
\begin{eqnarray}
    \text{BIC}=\chi^2_{\text{min}}+d\times\ln(N)
\end{eqnarray}
In this context, N denotes the number of data samples utilized in the MCMC analysis. Likewise, $\Delta\text{BIC}<2$ indicates strong evidence supporting the assumed theoretical model best, while a decrease in BIC between 2 and 6 suggests moderate support. By utilizing the aforementioned $\chi^2_{\text{min}}$ value, we obtained $\Delta\text{AIC}=1.408$ and $\Delta\text{BIC}=6.869$ where we have determined $\text{AIC}_{\text{Model}}=1654.988$, $\text{AIC}_{ \Lambda\text{CDM}}=1653.58$, $\text{BIC}_{\text{Model}}=1671.369$ and $\text{BIC}_{ \Lambda\text{CDM}}=1664.50$. Thus, it is evident from the $\Delta\text{AIC}$ and $\Delta\text{BIC}$ value that there is strong evidence in favor of our proposed Hubble function over the standard $\Lambda$CDM.

\subsection{Evolutionary Profile of Cosmological Parameters}
\justifying
In this section, we investigate the behavior of some cosmological parameters such as the deceleration parameter $q$, the jerk parameter $j$, and the snap parameter $s$, that plays a vital role to study the expansion phase of the universe.
\begin{figure}[H]
\includegraphics[width=6.1cm, height=4.3cm]{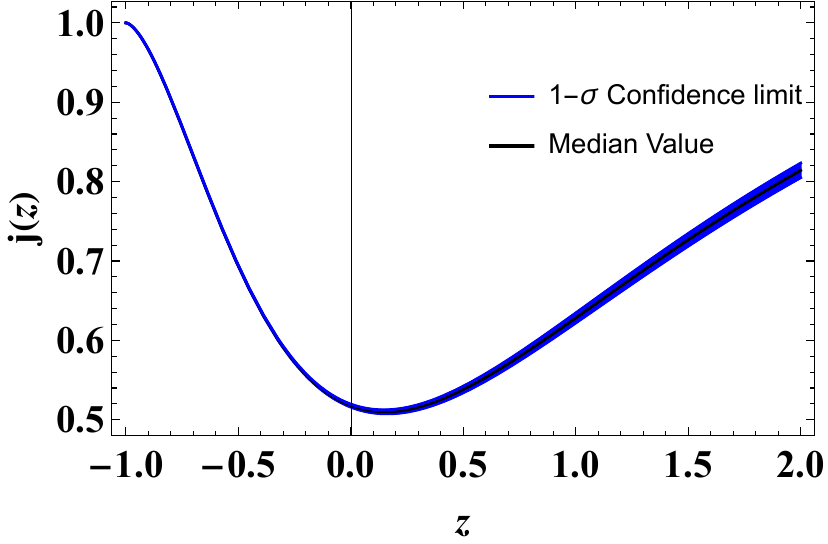}
\includegraphics[width=6.0cm, height=4.3cm]{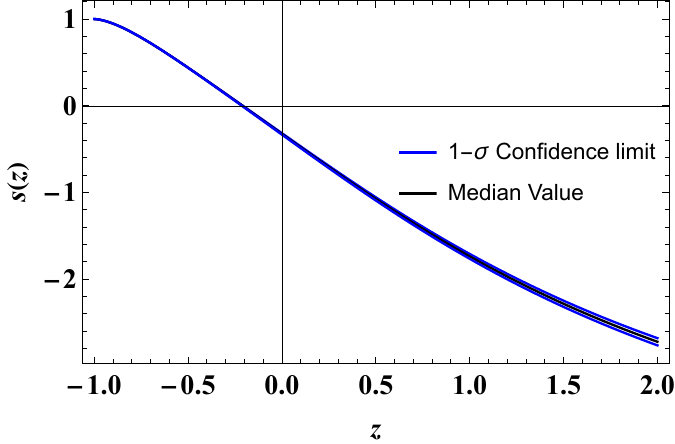}
\includegraphics[width=6.8cm, height=4.3cm]{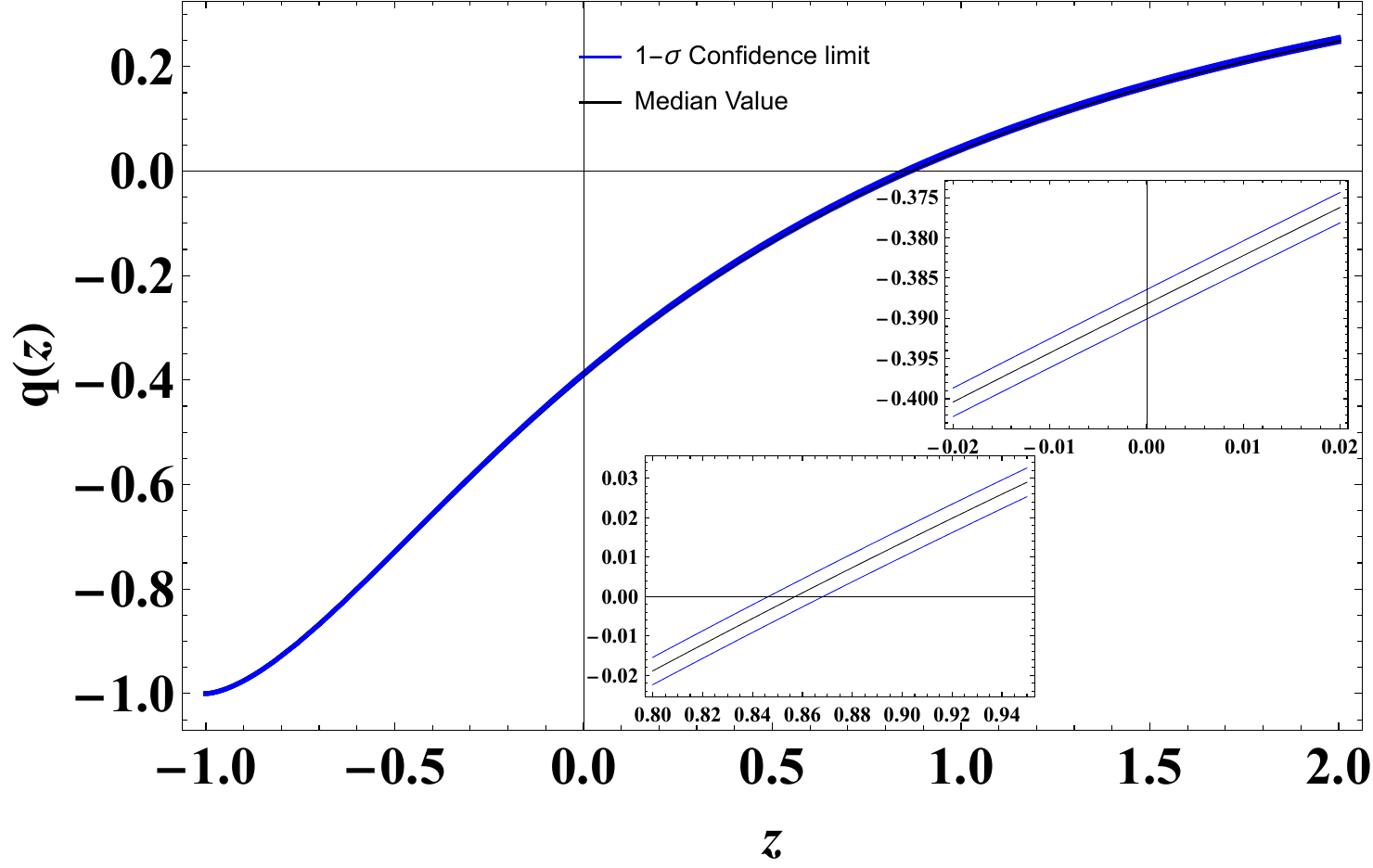}
\caption{Profile of the jerk, snap, and the deceleration parameter vs redshift.}\label{fig2} 
\end{figure}     

It is well known that the cosmographic parameters can be obtained as the coefficient of the Taylor series expansion of the scale factor with the present time as the center. The coefficient of terms $(t-t_0)^2$, $(t-t_0)^3$, and $(t-t_0)^4$ are defined as the deceleration parameter $q$, the jerk parameter $j$, and the snap parameter $s$, whereas the coefficient of the term $(t-t_0)$ known as the Hubble parameter. The explicit expression can be expressed as $q=-\frac{\ddot{a}}{aH^2}$, $j=\frac{\dddot{a}}{aH^3}$, and $s=\frac{\ddddot{a}}{aH^4}$ \cite{CS}. The evolutionary trajectories of these parameters corresponding to the proposed Hubble function have been presented in Fig.~\ref{fig2}. From the Fig.~\ref{fig2} it is evident that the proposed function predicts the de-Sitter type accelerated expansion phase at the late times via a transition epoch from the decelerated epoch to accelerated epoch in the recent past with the transition redshift $z_t=0.857 \pm 0.011$. The present value of all these parameters are $q_0=-0.388 \pm 0.002$, $j_0=0.517 \pm 0.002$, and $s_0=-0.325^{+0.008}_{-0.009}$ within $68 \%$ confidence limit. Note that $j(z) \equiv 1$, $s(z) \equiv 1$ in case of the $\Lambda$CDM model. The obtained present deceleration parameter value with transition redshift is quite consistent with the cosmological observations \cite{PL}.

\section{Cosmological $f(Q)$ models}\label{sec5}
\justifying
The choice of the function $f(Q)=Q$ presents an equivalent formulation to general relativity, and hence in the case, the theory is known as symmetric teleparallel equivalent to general relativity (STEGR). The correction to the STEGR case provides an interesting class of solutions applicable either in early or late-time cosmology \cite{13}. We consider some well-known corrections to the STEGR case that have been previously investigated in the coincident gauge formalism as follows,\\

\textbf{Model-I: $f(Q)=Q+\eta e^{\alpha Q}$} \cite{MS} 

\textbf{Model-II: $f(Q)=Q+\eta log{\alpha Q}$} \cite{KV}

\textbf{Model-III: $f(Q)=Q+\eta Q^\alpha$}  \cite{MS}

\textbf{Model-IV: $f(Q)=Q+\eta Q^{-1}$ }  \cite{a15}

\textbf{Model-V: $f(Q)=Q+\eta Q^2$ } \cite{a15}\\

For our further investigation, we have considered a non-constant function $\gamma(t)=-a^{-1}\dot{H}$, where $\dot{H}$ represents the time derivative of $H$. Note that, the choice $\gamma(t)=0$ reduces to coincident gauge formalism, whereas the case of constant $\gamma$ is the trivial one i.e. not much of the physical interest. Therefore we have assumed a non-constant choice of $\gamma$ function, however, one can construct the different choice of $\gamma(t)$ as done in \cite{GA}.

\subsection{Energy Conditions}
Energy conditions (ECs) are crucial for comprehending the universe's geodesics. These conditions can be derived from the established Raychaudhuri equations \cite{en1}. These conditions involve the contraction of timelike or null vector fields with the Einstein tensor and the energy-momentum tensor. These conditions help to define what is considered physically reasonable or not clearly. Therefore, it is beneficial to apply one or more energy conditions for studying a cosmological model \cite{en2,en3}. 
The mathematical expression for the strong energy condition (SEC) is $\rho+3p\geq 0$, whereas for the null energy condition (NEC) is $\rho+p\geq 0$. 
\begin{figure}[H]
\includegraphics[width=8.8cm, height=5.5cm]{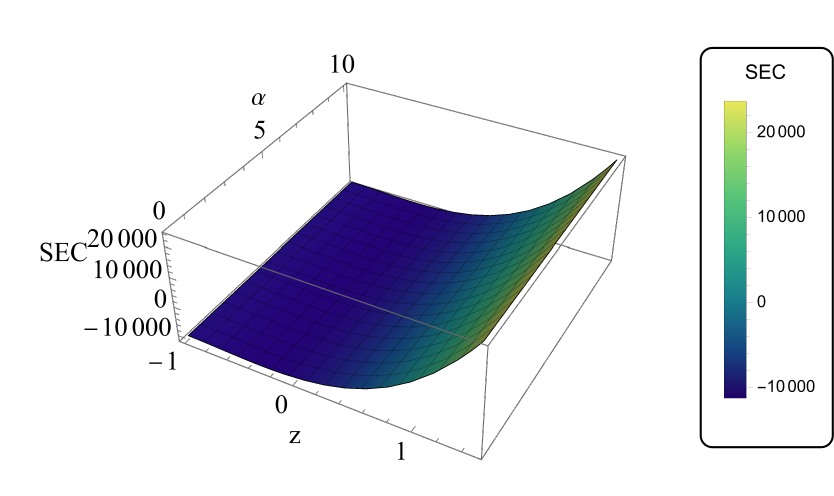}    
\includegraphics[width=9.0cm, height=5.5cm]{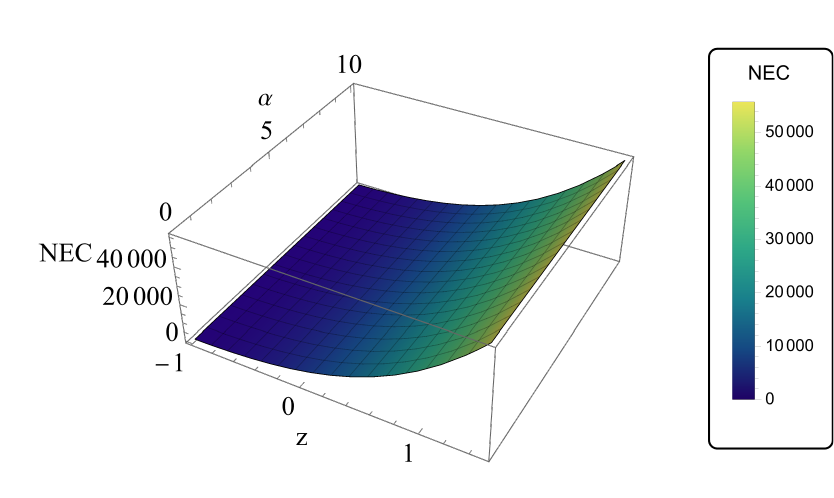}    
\caption{Profile of the SEC and the NEC vs redshift, corresponding to the Model-I with varying $\alpha$ and fixed $\eta=1$.}\label{M1_p} 
\end{figure}    

\begin{figure}[H]
\includegraphics[width=8.8cm, height=5.5cm]{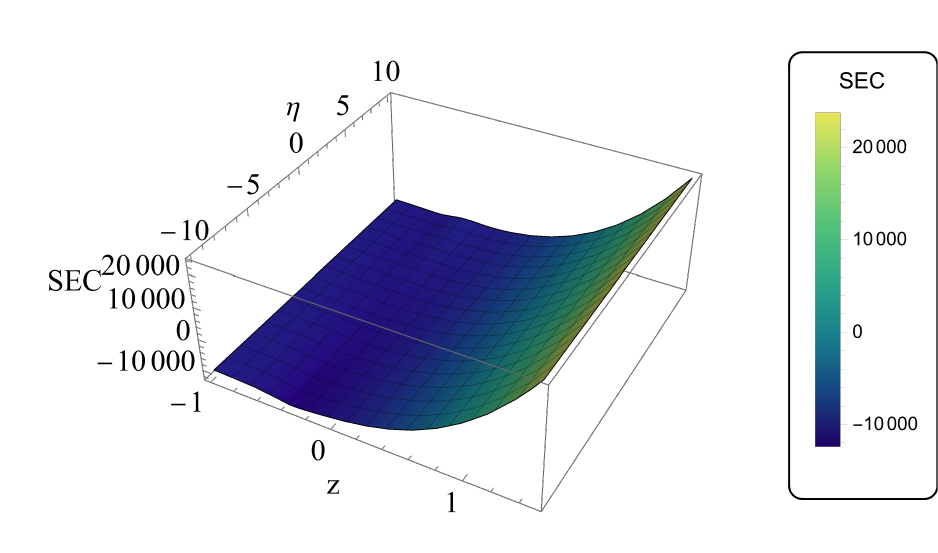}      
\includegraphics[width=9.0cm, height=5.5cm]{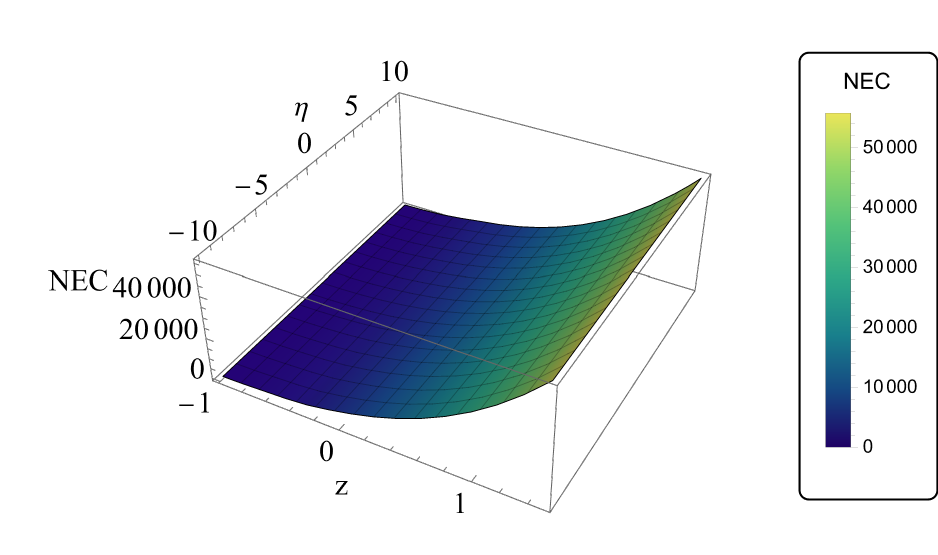}    
\caption{Profile of the SEC and the NEC vs redshift, corresponding to the Model-I with varying $\eta$ and fixed $\alpha=1$.}
\label{M1_c} 
\end{figure}        

\textbf{Model-I:} The Model-I is an exponential correction to the STEGR case. From Fig.~\ref{M1_p} and Fig.~\ref{M1_c}, it is clear that, in the entire range of redshift, the NEC is satisfied for both cases i.e, $\alpha \in [0,10]$ with $\eta=1$ and $\eta \in [-10,10]$ with $\alpha=1$. Further, the SEC is violated corresponding to both case in the recent past that favors the accelerating nature of expansion along with transition epoch in the recent past.\\

\begin{figure}[H]
\includegraphics[width=8.8cm, height=5.5cm]{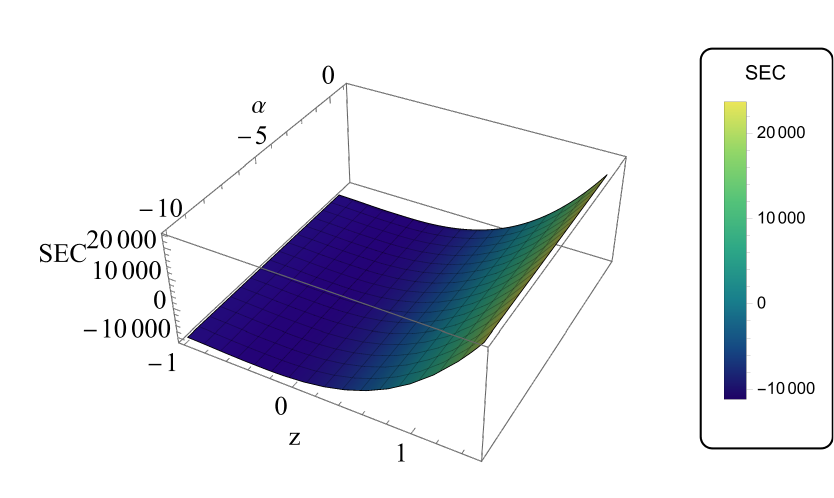}   
\includegraphics[width=9.0cm, height=5.5cm]{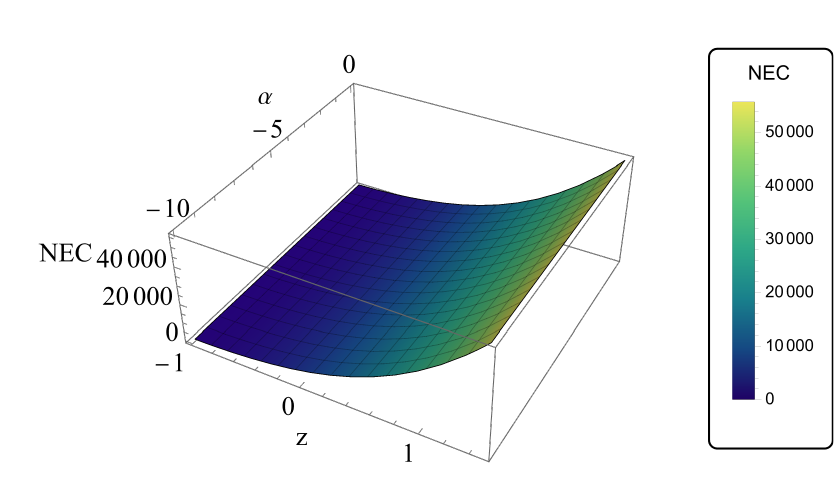}    
\caption{Profile of the SEC and the NEC vs redshift, corresponding to the Model-II with varying $\alpha$ and fixed $\eta=1$. }\label{M2_p} 
\end{figure}

\begin{figure}[H]
\includegraphics[width=8.8cm, height=5.5cm]{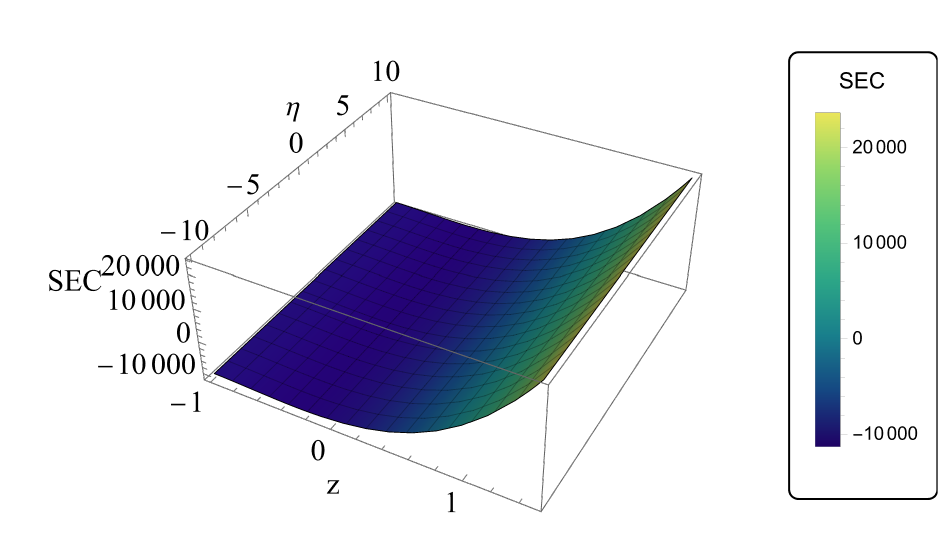}  
\includegraphics[width=9.0cm, height=5.5cm]{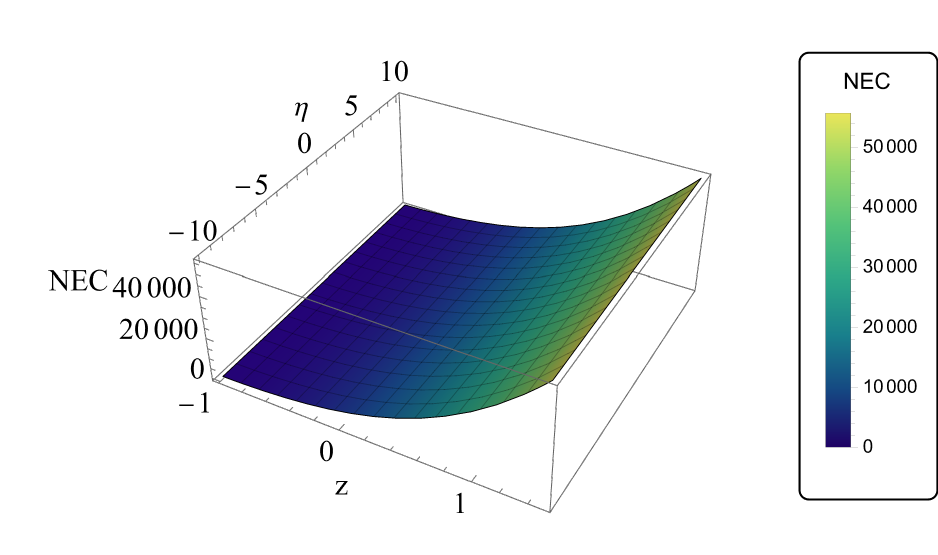}    
\caption{Profile of the SEC and the NEC vs redshift, corresponding to the Model-II with varying $\eta$ and fixed $\alpha=-1$.}\label{M2_c} 
\end{figure}

\textbf{Model-II:} The Model-II is an logarithmic correction to the STEGR case. From Fig.~\ref{M2_p} and Fig.~\ref{M2_c}, it is clear that, in the entire range of redshift, the NEC is satisfied for both cases i.e, $\alpha \in [-10,0]$ with $\eta=1$ and $\eta \in [-10,10]$ with $\alpha=-1$. Further, the SEC is violated corresponding to both case in the recent past that favors the accelerating nature of expansion along with transition epoch in the recent past.\\

\begin{figure}[H]
\includegraphics[width=8.8cm, height=5.5cm]{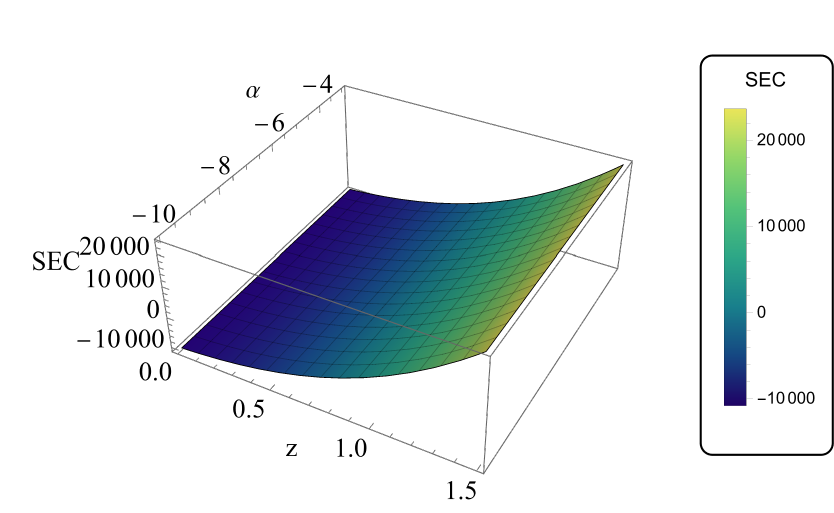}  
\includegraphics[width=9.0cm, height=5.5cm]{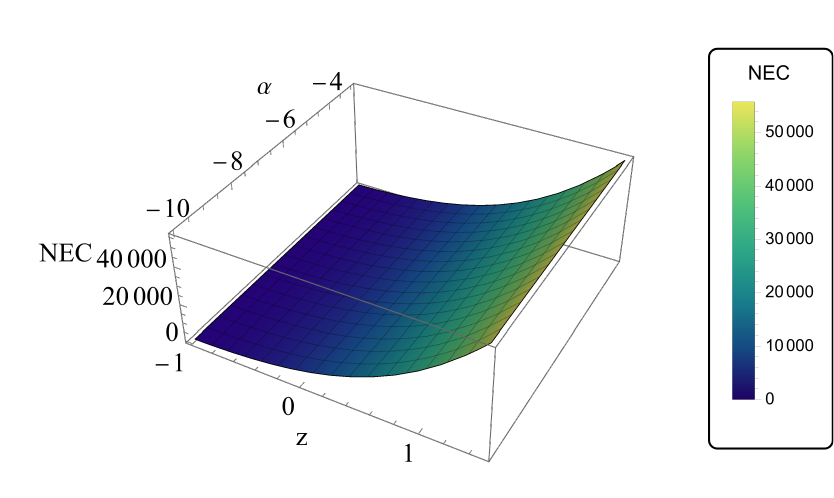} 
\caption{Profile of the SEC and the NEC vs redshift, corresponding to the Model-III with varying $\alpha$ and fixed $\eta=1$.}\label{M3_p} 
\end{figure}

\textbf{Model-III:} The Model-III is a power-law correction to the STEGR case. From Fig.~\ref{M3_p}, it is evident that, in the entire range of redshift, the NEC is satisfied for the parameter value $\alpha \in [-10,-4]$ with $\eta=1$. Further, the SEC is violated in the recent past that favors the accelerating nature of expansion along with transition epoch in the recent past. It is interesting to note that the same power-law correction model considered in \cite{MS} in the \textit{coincident gauge connection} favors the positive values of $\alpha$ with the negative value of coefficient $\eta$. \\

\begin{figure}[H]
\includegraphics[width=8.8cm, height=5.5cm]{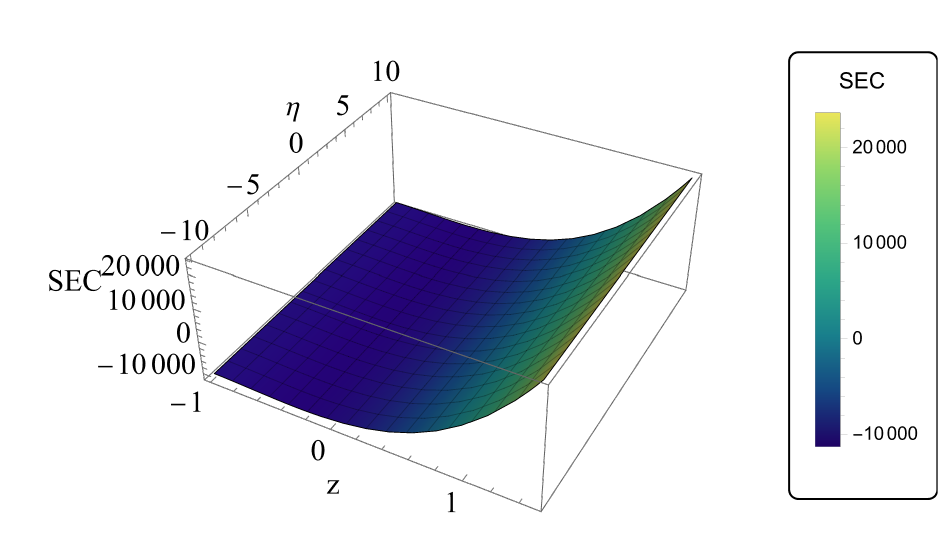} 
\includegraphics[width=9.0cm, height=5.5cm]{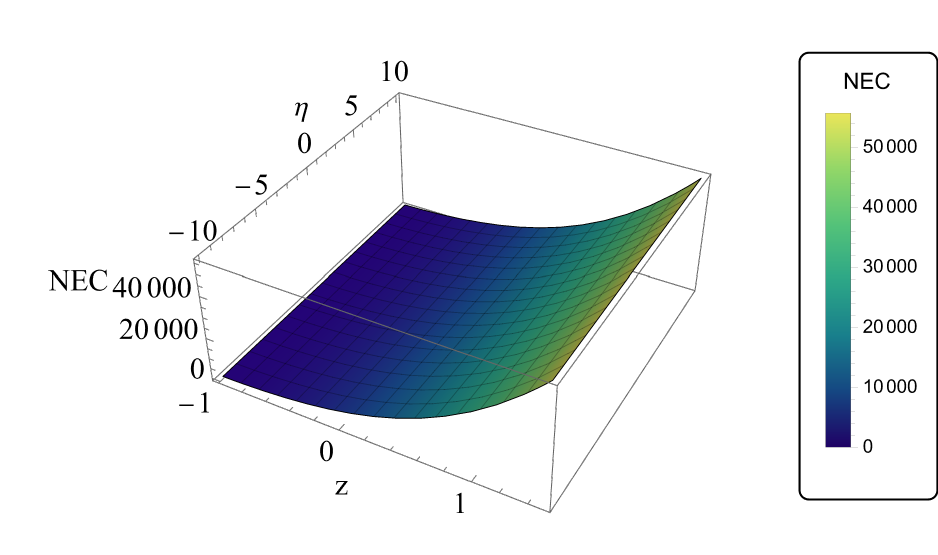} 
\caption{Profile of the SEC and the NEC vs redshift, corresponding to the Model-IV with varying $\eta$ and fixed $\alpha=-1$.}
\label{M4_p} 
\end{figure}

\textbf{Model-IV:} The Model-IV is a negative power-law correction to the STEGR case. A negative power correction to the STEGR case provides a correction to the late-time cosmology, where they can give rise to the dark energy  \cite{a15}. From Fig.~\ref{M4_p}, it is evident, in the entire range of redshift, the NEC is satisfied for the parameter value $\eta \in [-10,10]$ with $\alpha=-1$. Further, the SEC is violated in the recent past that favors the accelerating nature of expansion along with transition epoch in the recent past. \\

\begin{figure}[H]
\includegraphics[width=7.0cm, height=5.5cm]{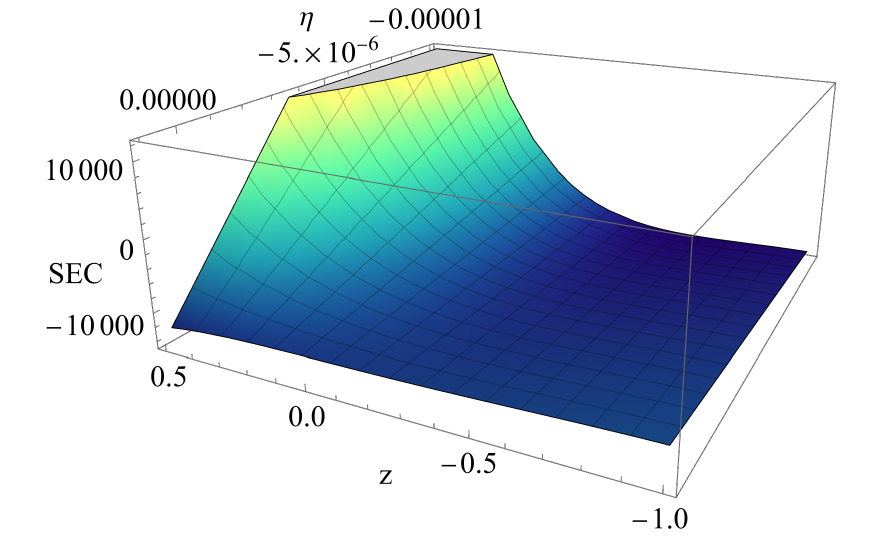} 
\includegraphics[width=1.6cm, height=5.0cm]{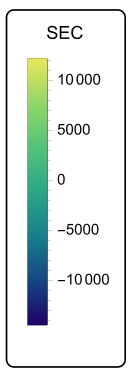} 
\includegraphics[width=7.0cm, height=5.5cm]{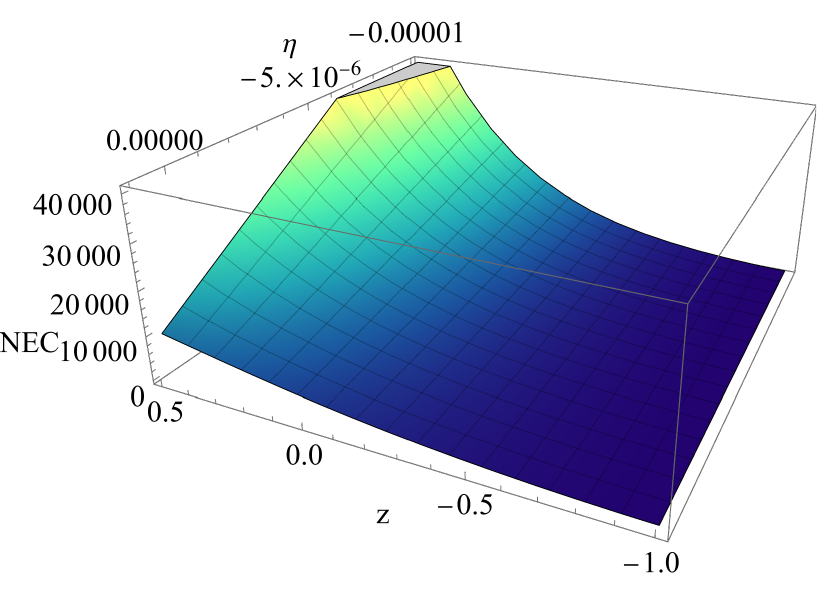} 
\includegraphics[width=1.5cm, height=5.0cm]{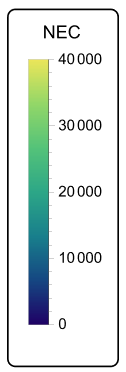}
\caption{Profile of the SEC and the NEC vs redshift, corresponding to the Model-V with varying $\eta$ and fixed $\alpha=2$.}
\label{M5_p} 
\end{figure} 

\textbf{Model-V:} The Model-V is a positive power-law correction to the STEGR case. A positive power correction to the STEGR case provides a correction to the early epochs of the universe with potential applications to inflationary solutions \cite{a15}. From Fig.~\ref{M5_p}, it is evident that the NEC is satisfied for the parameter value $\alpha=2$ and for a tiny value of $\eta$ nearly close to $0$. Further, the SEC is violated in the recent past that favors the accelerating nature of expansion along with transition epoch in the recent past. 

\subsection{Thermodynamical Stability}
\justifying
A rigorous criterion for validating a cosmological model is the square of the speed of sound $v_s^2$. For a model to be physically plausible, the square of the speed of sound must be less than the speed of light $c$. This relationship establishes the stability requirement for cosmological models. Therefore, if the condition  $0<v_s^2<c$ is satisfied, the model is considered physically plausible. These constraints enhance the suitability of this type of modeling, and various models featuring variable sound speed have been documented in the literature \cite{sos1,sos2}.
For this analysis, we consider the universe as an adiabatic system, implying no heat or mass exchange with its surroundings and thus maintaining zero entropy perturbation. Under these assumptions, the primary focus shifts to how pressure varies with energy density, which leads us to define the sound speed parameter by the following expression:.
\begin{eqnarray}
v_s^2=\frac{dp}{d\rho}
\end{eqnarray}

\begin{figure}[H]
\includegraphics[width=6.5cm, height=4.8cm]{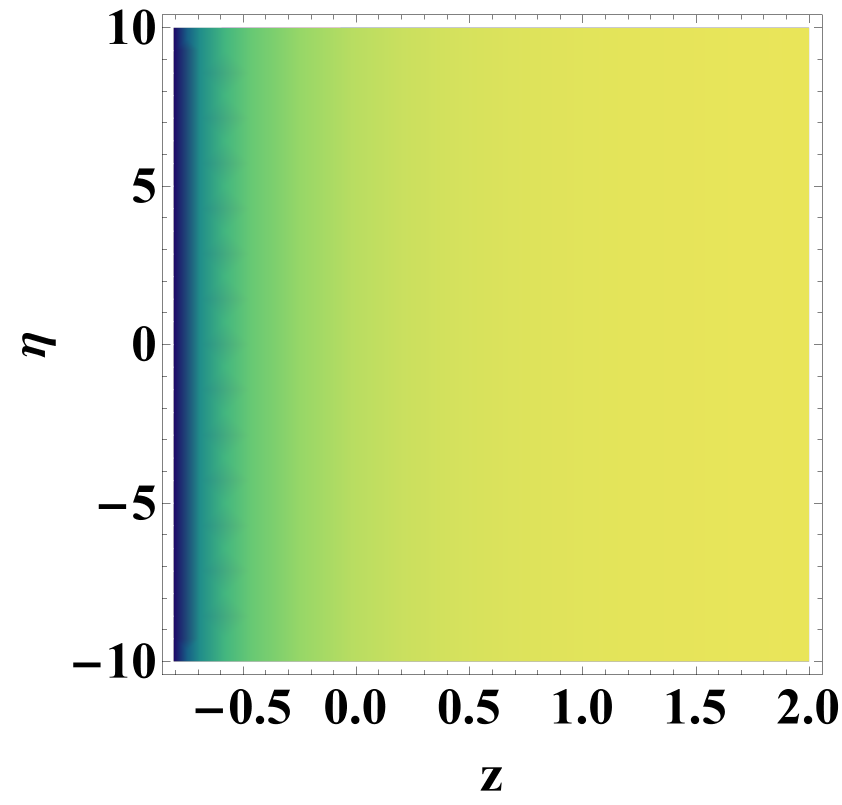}
\includegraphics[width=1.0cm, height=4.8cm]{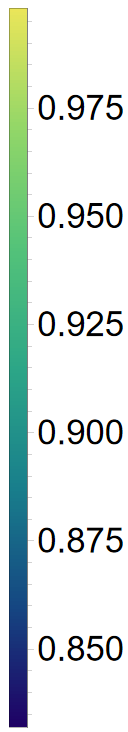}
\includegraphics[width=6.8cm, height=4.8cm]{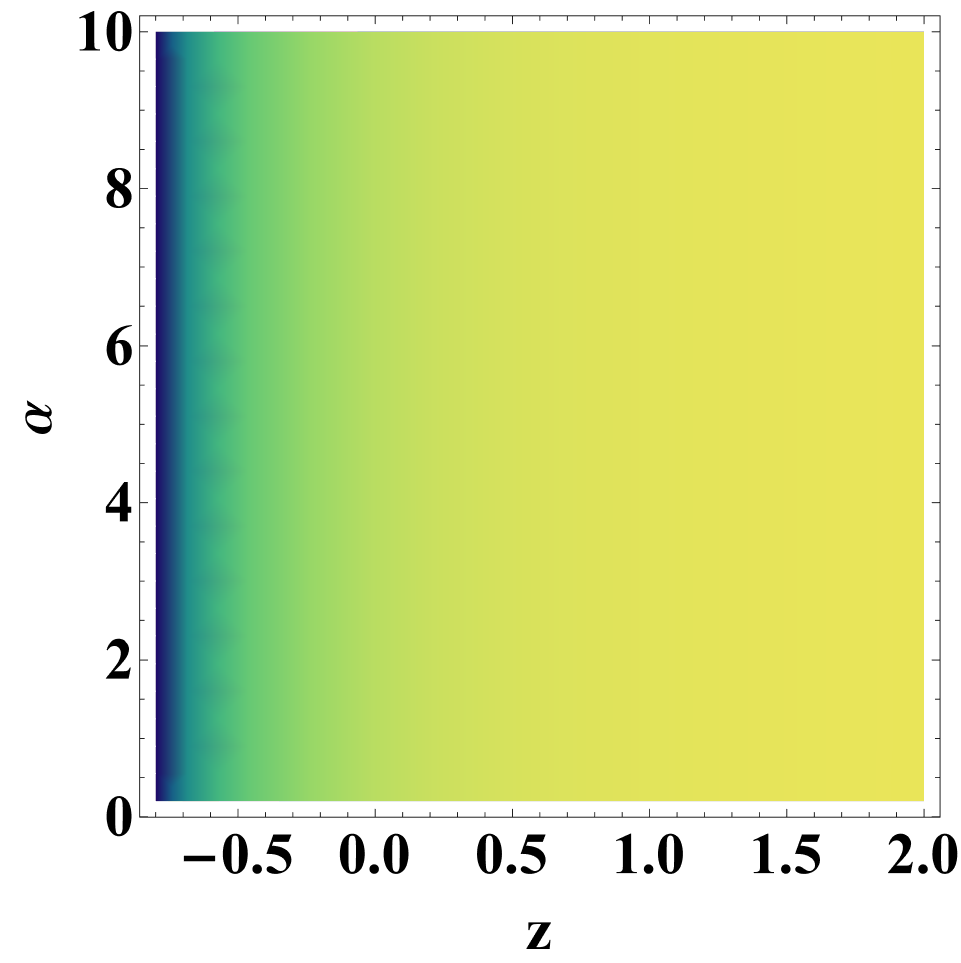}
\includegraphics[width=1.0cm, height=4.8cm]{m3stab2bar2.png}
\caption{Profile of the sound speed parameter vs z (redshift) for the Model-I corresponding to the case varying $\eta$ with $\alpha=1$ (left panel) and the varying $\alpha$ with $\eta=1$ (right panel).}\label{stab1} 
\end{figure}
From the Fig.~\ref{stab1}, one can observe that the sound speed parameter for the Model-I lies between $0$ and $1$ corresponding to both cases i.e, $\eta \in [-10,10]$ with $\alpha=1$ (left panel) and $\alpha \in [0,10]$ with $\eta=1$ (right panel). Thus the assumed exponential $f(Q)$ correction model show stable behavior in the same range obtained from the energy conditions.

\begin{figure}[H]
\includegraphics[width=6.5cm, height=4.8cm]{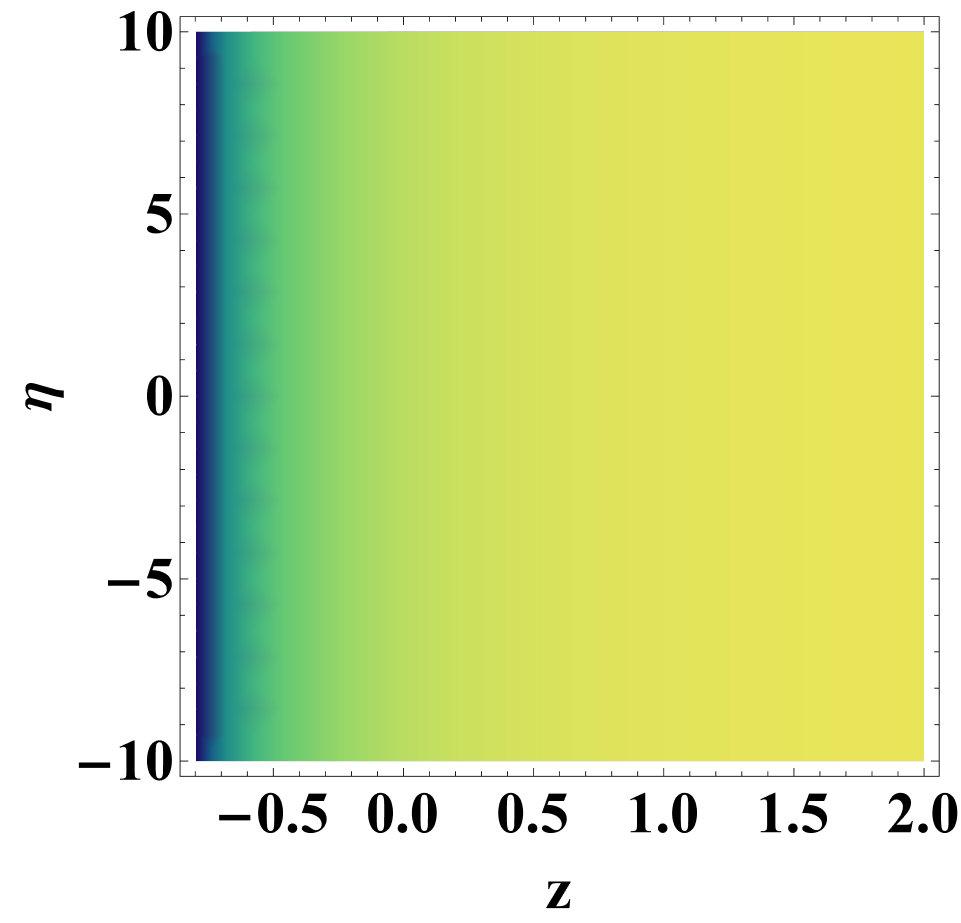}
\includegraphics[width=1.0cm, height=4.8cm]{m3stab2bar2.png}
\includegraphics[width=6.5cm, height=4.8cm]{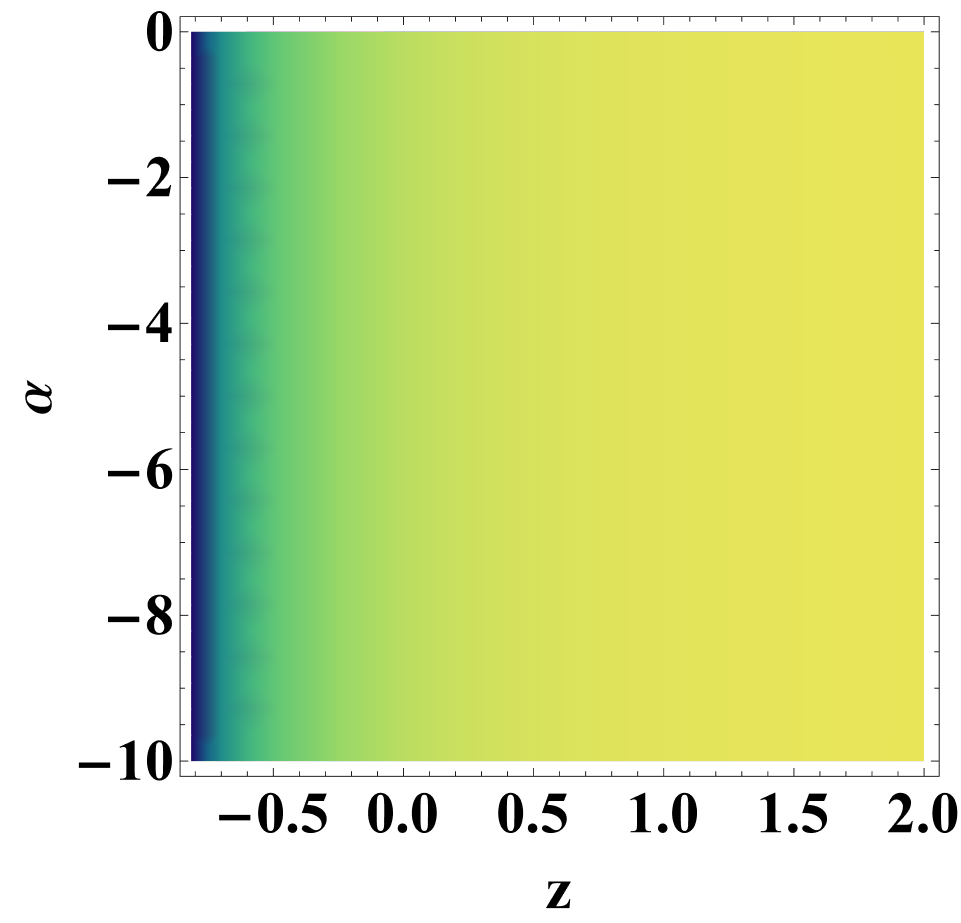}
\includegraphics[width=1.0cm, height=4.8cm]{m3stab2bar2.png}
\caption{Profile of the sound speed parameter vs z (redshift) for the Model-II corresponding to the case varying $\eta$ with $\alpha=-1$ (left panel) and the varying $\alpha$ with $\eta=1$ (right panel).}\label{stab2} 
\end{figure}
From the Fig.~\ref{stab2}, it is evident that the sound speed parameter for the Model-II lies between $0$ and $1$ corresponding to both cases i.e, $\eta \in [-10,10]$ with $\alpha=-1$ (left panel) and $\alpha \in [-10,0]$ with $\eta=1$ (right panel). Thus the assumed logarithmic $f(Q)$ correction model show stable behavior in the same range obtained from the energy conditions.

\begin{figure}[H]
\includegraphics[width=5.1cm, height=4.1cm]{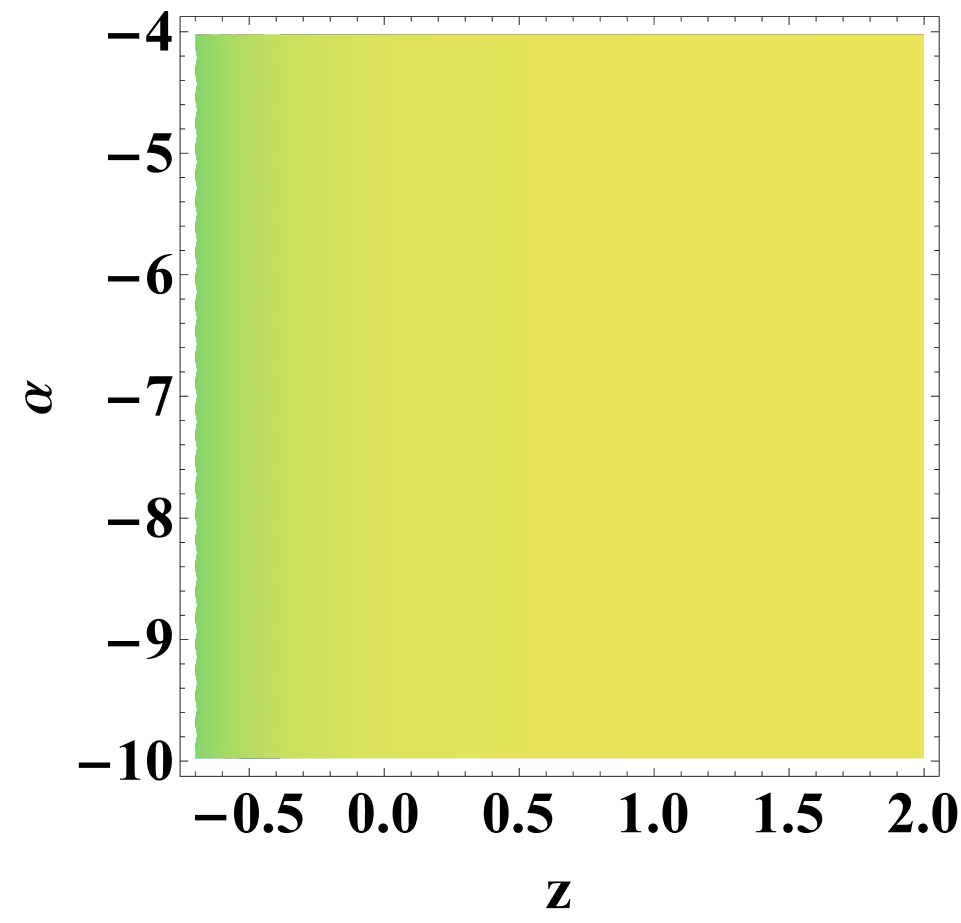}
\includegraphics[width=0.68cm, height=4.1cm]{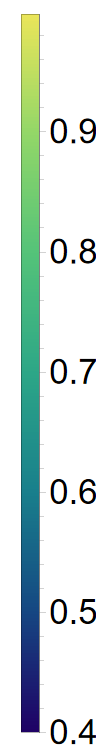}
\includegraphics[width=5.3cm, height=4.1cm]{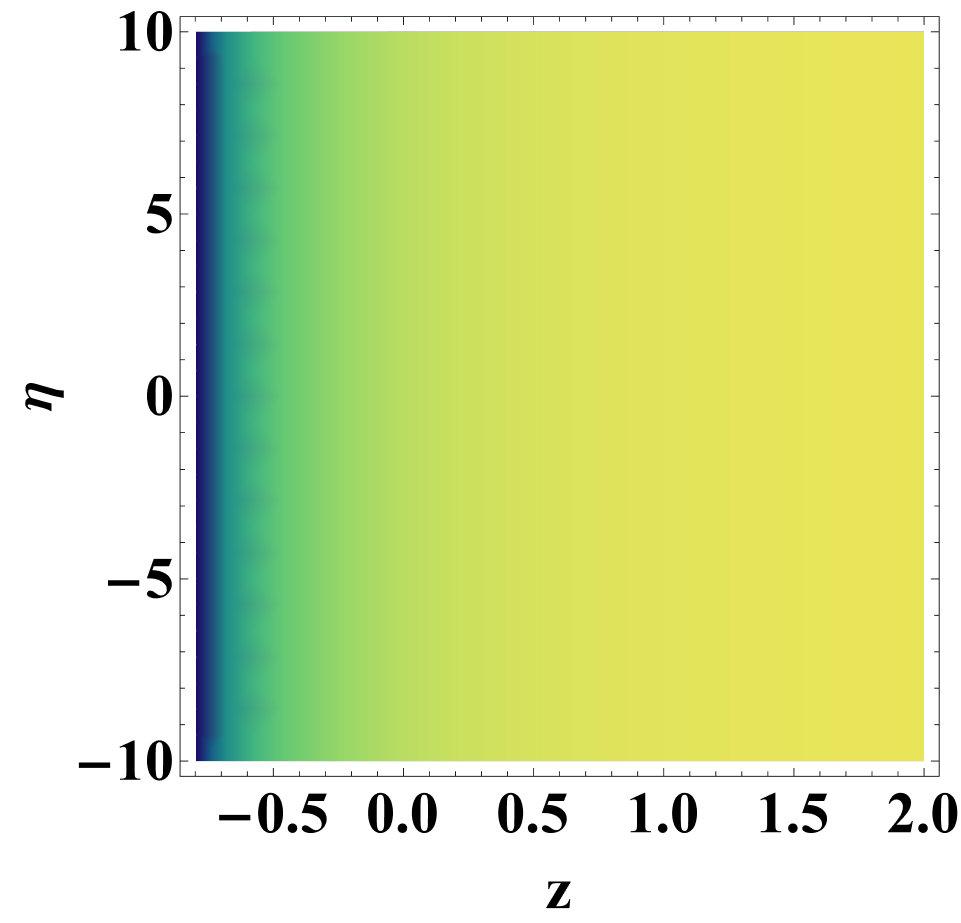}
\includegraphics[width=0.670cm, height=4.1cm]{m3stab2bar2.png}
\includegraphics[width=5.2cm, height=4.1cm]{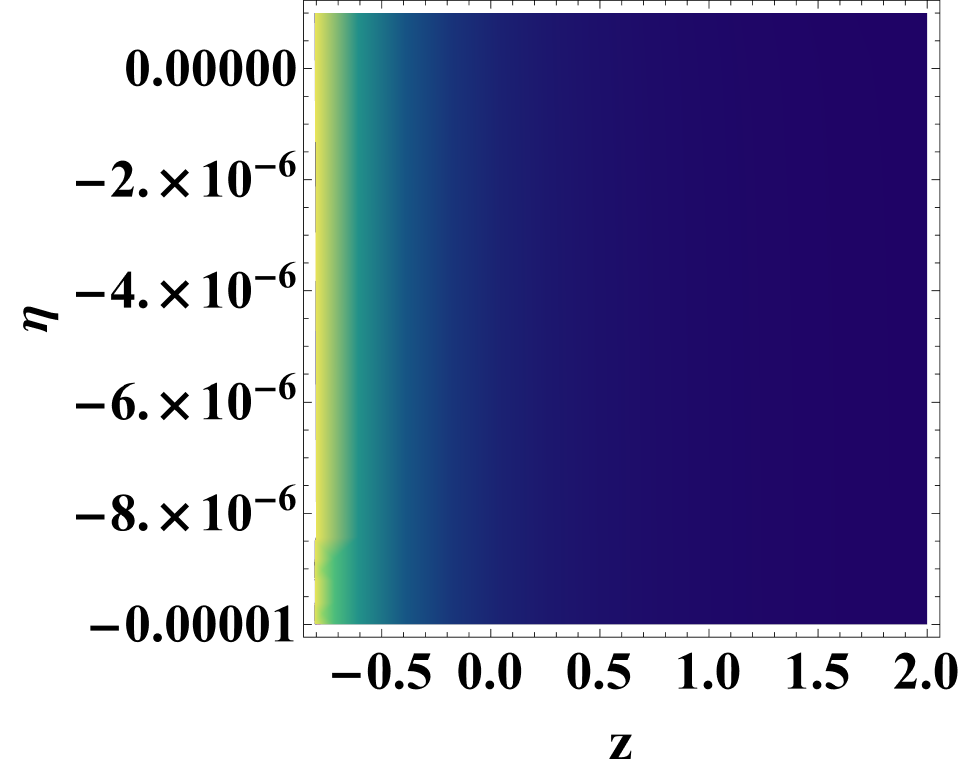}
\includegraphics[width=0.67cm, height=4.1cm]{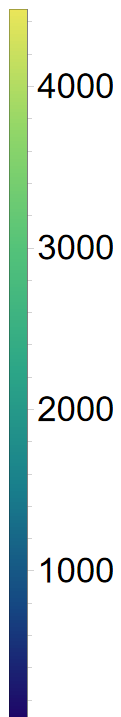}
\caption{Profile of the sound speed parameter vs z (redshift) for the Model-III with varying $\alpha$ and $\eta=1$ (left panel), Model-IV with varying $\eta$ and $\alpha=-1$ (middle panel), and the Model-V with varying $\eta$ and $\alpha=2$ (right panel).}\label{stab3} 
\end{figure}
From the Fig.~\ref{stab3} (left panel), we find that the sound speed parameter for the Model-III lies between $0$ and $1$ corresponding to the parameter choice $\alpha \in [-10,-4]$ with $\eta=1$ (left panel). Thus the considered power-law $f(Q)$ correction model show stable behavior in the same range obtained from the energy conditions. Further, we have assumed two particular cases of the power-law model with both negative and positive power. Both assumed model have significantly utilized in the case of coincident gauge choice and found to be physically viable. The Model-IV is a negative power-law correction that can provides a correction to the late-time cosmology, where they can give rise to the dark energy, whereas the Model-V is a positive power-law correction that can provides a correction to the early epochs of the universe with potential applications to inflationary solutions \cite{a15}. In this investigation, we observe that the sound speed parameter for the Model-IV lies between $0$ and $1$ corresponding to the parameter choice $\eta \in [-10,10]$ with $\alpha=-1$ (see Fig.~\ref{stab3} middle panel), whereas for the Model-V, it does not lies between $0$ and $1$ corresponding to the parameter choice obtained from energy condition constraints (see Fig.~\ref{stab3} right panel).  Thus the considered negative power-law $f(Q)$ correction model show stable behavior, whereas the positive power-law $f(Q)$ correction model show unstable behavior.

\section{Conclusion}\label{sec6}
\justifying
In the symmetric teleparallel equivalent of general relativity, gravity is characterized by non-metricity in an environment devoid of both curvature and torsion. We have investigated a natural extension of this framework, the modified $f(Q)$ theory, which has been demonstrated to successfully explain the late-time acceleration of the universe without the need to assume dark energy. Our current study focuses specifically on the cosmological aspects of this theory. It has been observed that researchers have predominantly used the coincident gauge formulation (assuming a vanishing affine connection) when applying $f(Q)$ theory to a spatially flat FLRW spacetime. This choice of gauge has a significant drawback, as the resulting Friedmann like equations are identical to those found in the well-studied torsion-based $f(T)$ theory. As a result, researchers have unintentionally been reproducing known results under a different gravity.

In this work, we have therefore utilized a new $f(Q)$ theory dynamics utilizing a non-vanishing affine connection involving a function $\gamma$, resulting in Friedmann equations that are entirely distinct from those of $f(T)$ theory. \textbf{The basic mathematical construction of $f(Q)$ gravity from a non-vanishing affine connection involves an arbitrary function of time $\gamma(t)$. There are three admissible connections. Some of the non-zero components are common, whereas the way a free function of time $\gamma(t)$ enters in some of their other components give rise to three inequivalent class of flat $f(Q)$ cosmologies arising due to these three different connection \cite{ab2}. As this function is arbitrary, one can investigate the different class of this function, as done in \cite{GA}.} We have considered a non-constant function $\gamma(t)=-a^{-1}\dot{H}$, where $\dot{H}$ represents the time derivative of $H$. Note that, the choice $\gamma(t)=0$ reduces to coincident gauge formalism, whereas the case of constant $\gamma$ is the trivial one i.e. not much of the physical interest. Therefore we have assumed a non-constant choice of $\gamma$ function, however, one can construct the different choice of $\gamma(t)$ as done in \cite{GA}. In addition, we have proposed a new parameterization of the Hubble function in the equation (\ref{NH}). As several previous parameterization schemes \cite{sib1,sib2,sib3} leads to a discrepancy from the standard $\Lambda$CDM and hence cannot consistently describe the present deceleration parameter value, transition redshift, and the late time de-Sitter limit as obtained in standard $\Lambda$CDM model. This motivated us to consider a new parameterization function that can describe the aforementioned cosmological epochs with observational compatibility.

We carried out a statistical analysis to evaluate the predictions of the assumed Hubble function parameterization by imposing constraints on the free parameters. We utilized Bayesian statistical analysis to estimate the posterior probability by employing the likelihood function and the Markov Chain Monte Carlo (MCMC) sampling technique by invoking the CC, Pantheon+SH0ES, and the BAO data samples for our analysis. We obtained the free parameter constraints as $H_0=68 \pm 0.094 \: km/s/Mpc$, $\beta=42^{+0.04}_{-0.041}$, and $n=1.6 \pm 0.0036$ within $68 \%$ confidence limit. The corresponding contour plot describing the correlation between the free parameters within the $1\sigma-3\sigma$ confidence interval is presented in Fig.~\ref{fig1}. To determine the reliability of MCMC analysis we have conducted the AIC and BIC statistical evaluations. We have determined $\Delta AIC=1.408$ and $\Delta BIC= 6.869$ which supports a shred of strong evidence in favor of our proposed Hubble function over the standard $\Lambda \text{CDM}$.

In addition, we analyzed some cosmographic parameters like $q(z)$, $j(z)$ and $s(z)$. The evolutionary paths of these parameters, corresponding to the proposed Hubble function, are illustrated in Fig.\ref{fig2}. As seen in Fig.\ref{fig2}, the proposed function forecasts a de-Sitter type accelerated expansion phase at late times, following a transition from a decelerated epoch to an accelerated epoch in the recent past, with the transition redshift $z_t=0.857 \pm 0.011$. The current values of these parameters are $q_0=-0.388 \pm 0.002$, $j_0=0.517 \pm 0.002$, and $s_0=-0.325^{+0.008}_{-0.009}$ within $68 \%$ confidence limit. It is noteworthy that for the $\Lambda$CDM model, $j(z) \equiv 1$ and $s(z) \equiv 1$. The obtained present deceleration parameter value, along with the transition redshift, aligns well with cosmological observations \cite{PL}.

Further, we have considered some well-known corrections to the STEGR case such as an exponentital $f(Q)$ correction, logarithmic $f(Q)$ correction, and a power-law $f(Q)$ correction, that have been previously investigated in the coincident gauge formalism \cite{MS,KV,a15}. We obtained constraints on free parameters of the considered $f(Q)$ correction models via energy conditions. For the Model-I (see Fig.~\ref{M1_p} and Fig.~\ref{M1_c}) the NEC is satisfied and the SEC is violated corresponding to both cases i.e, $\alpha \in [0,10]$ with $\eta=1$ and $\eta \in [-10,10]$ with $\alpha=1$. For the Model-II (see Fig.~\ref{M2_p} and Fig.~\ref{M2_c}) the NEC is satisfied and the SEC is violated corresponding to both cases i.e, $\alpha \in [-10,0]$ with $\eta=1$ and $\eta \in [-10,10]$ with $\alpha=-1$. For the Model-III (see Fig.~\ref{M3_p}) the NEC is satisfied and the SEC is violated corresponding to the parameter value $\alpha \in [-10,-4]$ with $\eta=1$. It is interesting to note that the same power-law correction model (i.e. Model-III) considered in \cite{MS} in the \textit{coincident gauge connection} favors the positive values of $\alpha$ with the negative value of coefficient $\eta$. For the Model-IV (see Fig.~\ref{M4_p}) the NEC is satisfied and the SEC is violated, corresponding to the parameter value $\eta \in [-10,10]$ with $\alpha=-1$. Lastly, for the Model-V (see Fig.~\ref{M5_p}) it is evident that the NEC is satisfied and the SEC is violated for the parameter value $\alpha=2$ and for a tiny value of $\eta$ nearly close to $0$.

Finally, to test the physical plausibility of the assumed $f(Q)$ models we conducted the thermodynamical stability analysis via the sound speed parameter. We found that the assumed exponential and logarithmic $f(Q)$ corrections in Model-I and Model-II respectively are thermodynamically stable in the parameter range obtained from the energy condition constraints (see Fig.~\ref{stab1} and Fig.~\ref{stab2} ). Moreover, the considered power-law $f(Q)$ correction in Model-III show stable behavior in the same range obtained from the energy conditions (see Fig.~\ref{stab3} left panel). Further, we have assumed two particular cases of the power-law model with both negative and positive power. Both assumed models have been significantly utilized in the case of coincident gauge choice and found to be physically viable. The Model-IV is a negative power-law correction that can provides a correction to the late-time cosmology, where they can give rise to dark energy, whereas the Model-V is a positive power-law correction that can provides a correction to the early epochs of the universe with potential applications to inflationary solutions \cite{a15}. In this investigation, we found that the considered negative power-law $f(Q)$ correction model shows stable behavior (see Fig.~\ref{stab3} middle panel), whereas the positive power-law $f(Q)$ correction model shows unstable behavior (see Fig.~\ref{stab3} right panel). \textbf{Further, the investigation \cite{PQ} shows that two branches of the $f(Q)$ class of models exhibit reduced linear spectra, signalling they are infinitely strongly coupled. For the remaining branch they unveil the presence of seven gravitational degrees of freedom and at least one of them is ghost. Their results show the non-viability of all the spatially flat $f(Q)$ cosmologies. However, despite not being suitable for physical applications, this framework does
possess a series of compelling theoretical features that can be interesting to explore.}

\section*{Acknowledgement}
SP \& PKS  acknowledges the National Board for Higher Mathematics (NBHM) under the Department of Atomic Energy (DAE), Govt. of India for financial support to carry out the Research project No.: 02011/3/2022 NBHM(R.P.)/R \& D II/2152 Dt.14.02.2022. RS acknowledges UGC, New Delhi, India for providing Senior Research Fellowship (UGC-Ref. No.: 191620096030).

\end{document}